\documentclass[12pt]{article}
\usepackage{graphics,epsfig,rotating}

\begin{document}
\begin{center}
\large{ \bf{
           Statistical evolution of fragment isospin in nuclear 
                    multifragmentation.}}
\end{center} 

\begin{center}
\large{
         A.S.~Botvina}
\end{center}
\normalsize
\begin{center}
 
 Gesellschaft f\"ur Schwerionenforschung, D-64291 Darmstadt, Germany\\ 
 and Institute for Nuclear Research, 117312 Moscow, Russia
\end{center}

\begin{abstract}
Two instructive effects concerning fragment production at 
disintegration of finite nuclei are predicted with the statistical 
multifragmentation model: (1) a concentration of neutrons in 
intermediate mass fragments during the phase transition, (2) the break 
of the spatial symmetry of the fragment's isospin distribution, as well 
as of the symmetry of the fragment's emission from a statistical source, 
induced by the external Coulomb field. 
\end{abstract}

\hspace{0.5cm}

The knowledge of the isotope composition of fragments produced in nuclear 
multifragmentation 
can help in resolving the important problems: 
Do the fragments keep the memory of the initial dynamical stage or 
are they produced statistically? 
How does the isospin influence disintegration of finite nuclei and what is the 
difference to the case of nuclear matter? What is the 
isospin dependence of the nuclear equation of state? Generally, 
this study addresses an intriguing interdisciplinary problem of 
the phase transition in a finite-size two-component system (i.e. in a 
nucleus consisting of neutrons and protons), that is instructive 
for all fields dealing with finite systems. 
The problem was investigated within the statistical multifragmentation 
model (SMM) \cite{PR95}, which is successfully used for explanation of 
experimental data. A new Markov chain method of partition generation 
was incorporated in the 
model \cite{mc_prc}, that allows for considering the 
multifragmentation process on a solid microcanonical basis. The presented 
results reflect statistical properties of the fragment production and can be 
used for identification of the phenomenon.



\begin{figure}[!ht]
\includegraphics[height=13cm]{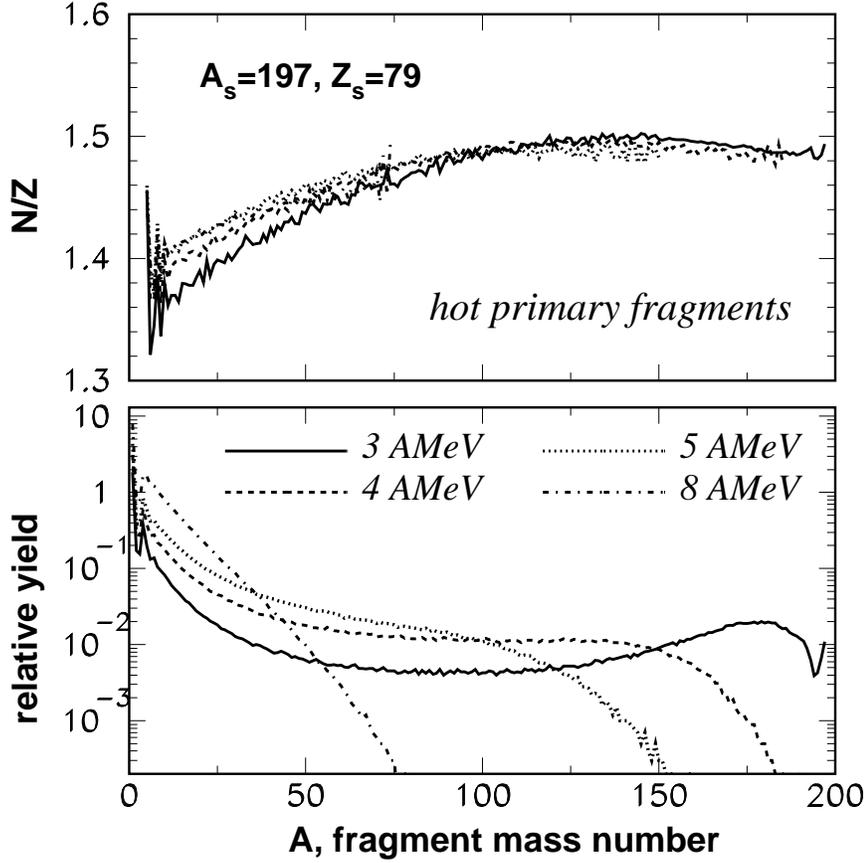}
\caption{The neutron-to-proton ratio N/Z and relative yield of 
hot primary fragments produced after break-up of Au nuclei at 
different excitation energies: 3 (solid lines), 4 (dashed lines), 
5 (dotted lines) and 8 (dot-dashed lines) MeV per nucleon.} 
\end{figure}

Presently, one of the extensively discussed topics is the isospin 
fractionation at disintegration of excited nuclei \cite{msu}. 
Fig.~1 shows mass distributions and neutron-to-proton ratios (N/Z) 
of the fragments produced after multifragmentation of a Au source 
(mass number $A_s$=197, charge $Z_s$=79), the calculations were performed 
at the standard SMM parameters \cite{PR95}. One can see a general statistical 
trend: the N/Z ratio of hot primary fragments increases with their mass 
numbers. This is a consequence of the interplay between 
the Coulomb and symmetry energy contributions to the binding energy of 
fragments \cite{PR95}. This trend persists up to $A \leq A_s/2$, while at 
larger $A$ the finite-size effects due to the mass and charge 
conservation prevail. In Fig.~1 
one can also see the evolution of the N/Z ratio and mass distribution of 
fragments in the excitation energy range $E_s^*$=3--8 MeV/nucleon. 
This energy range is usually associated with a liquid-gas type phase 
transition in finite nuclei, where 
the fragment mass distribution evolves from the U--shape, 
at the multifragmentation threshold $E_s^* \sim 3$ MeV/nucleon, 
to an exponential fall at high energies. During this evolution the 
temperature reaches a "plateau" and is nearly constant \cite{PR95}. 
As the energy increases the N/Z ratio of primary 
intermediate mass fragments (IMF, charges Z=3--20) increases, too. 
The reason is that the heaviest neutron-rich fragments are destroyed at 
increasing excitation energy, and some of their neutrons are bound in the 
IMFs, since the number of free neutrons is still small at this stage. 
Simultaneously, the N/Z ratio of the heaviest fragments decreases slightly. 
At very high excitation energy ($E_s^* > 8$ MeV/nucleon) the N/Z ratio of 
IMFs does not rise anymore but drops because no heavier 
fragments are left and the number of free neutrons increases rapidly, 
together with the temperature. 
This isospin evolution shows how the isospin 
fractionation phenomenon predicted for nuclear matter \cite{mueller} 
actually shows up in finite nuclear systems. In the region 
associated with the phase transition in neutron-rich nuclear systems we 
expect rather increasing neutron content of IMFs than increasing the number 
of free neutrons. Such a mechanism is consistent with recent experimental 
data \cite{milazzo}.

In peripheral nucleus--nucleus collisions at projectile energies of 
10--100 MeV/nucleon the break-up of highly excited projectile-like nuclei is 
fast (the characteristic time is around 100 fm/c) and happens in 
the vicinity of a target-like residue. In this case the Coulomb field of 
the target residue can influence the fragmentation of the projectile source 
and break the symmetry of the phase space population which exists 
for an isolated statistical source. 
This leads to spatial asymmetry of the fragment emission: small fragments 
are preferably emitted to the side of the target \cite{botvina99}. 
SMM calculations were performed for the Au projectile source 
which was placed at a fixed distance (20 fm) from another Au source 
simulating the target residue. This 
distance was obtained under the assumption that the break-up happens at 
$\sim$100 fm/c after a peripheral collision of a 35 $A\cdot$MeV Au projectile 
with a Au target. 

\begin{figure}
\includegraphics[width=14cm]{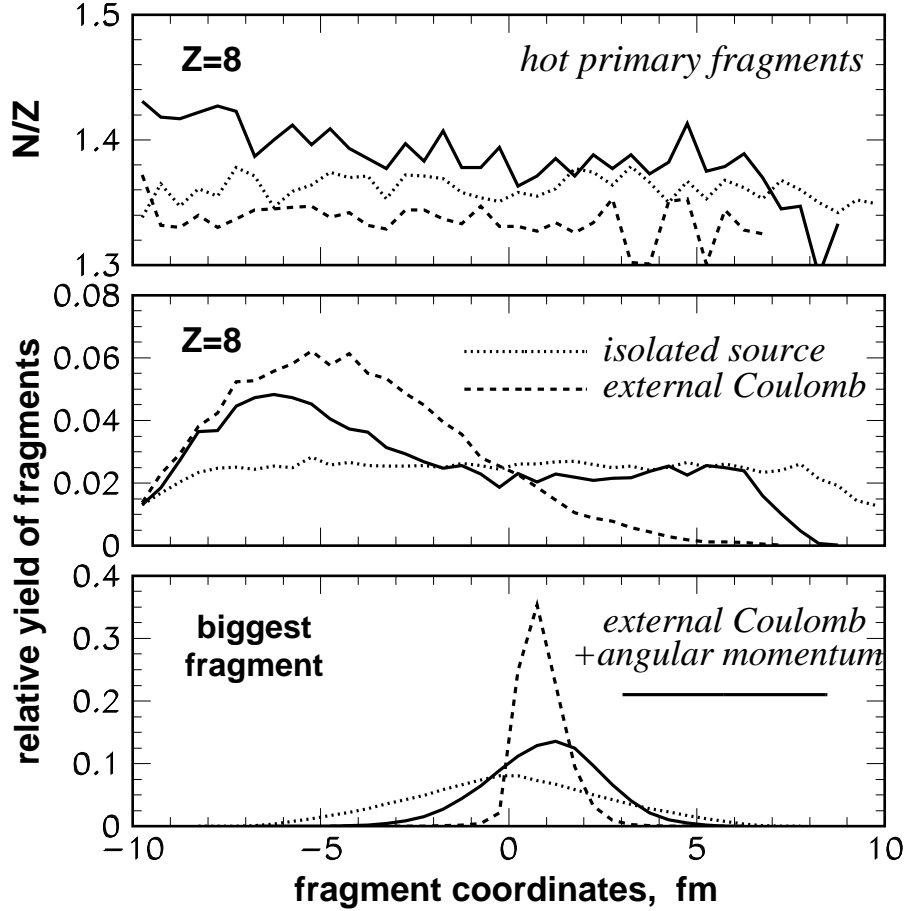}
\caption{Freeze-out coordinate distributions of the neutron-to-proton 
ratio N/Z of primary fragments with Z=8 (top panel) and relative yields of 
the primary Z=8 and biggest fragments (middle and bottom panels) produced 
at break-up of a Au source at an excitation energy of 3 MeV/nucleon. 
The freeze-out density is $\rho_s$=$\rho_0$/6 ($\rho_0\approx$ 0.15 fm$^{-3}$ 
is the normal nuclear density). 
The second Au nucleus is placed at -20 {\em fm} from the geometrical 
center of the freeze-out volume. 
Dotted lines: the isolated Au source, dashed lines: Coulomb interaction  
with the second Au nucleus is included, solid lines: angular momentum of 
150 $\hbar$ is included as well.}
\end{figure}

Fig.~2 shows the spatial distributions of yields and N/Z ratios 
of hot primary IMFs 
with $Z$=8 and the biggest fragments in the freeze-out volume along the 
axis connecting the projectile and target sources. It is seen that in the case 
of a single isolated source all distributions are symmetric with respect to 
the center of mass of the source. In the case of an 
external Coulomb field induced by the target source, the IMFs are mainly 
located at the target side while the biggest fragments are shifted to the 
opposite direction. Such positioning of fragments minimizes the Coulomb 
energy of the target-projectile system. If we take into account an angular 
momentum possibly transfered to the projectile source during the collision 
the N/Z ratio of the IMFs increases 
considerably and becomes larger for the IMFs which are closer to the target. 
The reason is an interplay between the Coulomb and rotational energy. 
An angular momentum favors the emission of IMFs with larger mass numbers, 
providing a larger moment of inertia, in oder to minimize the rotational 
energy and maximize the entropy. On the other side the Coulomb interaction, 
depending also on the fragment distance to the target, prevents the emission 
of IMFs with large charges. As a result of the interplay of these two factors 
we obtain neutron-rich fragments. The subsequent Coulomb propagation pushes 
the IMFs in the direction of the target providing predominant population 
of the midrapidity kinematic region by neutron-rich IMFs. 
This mechanism should 
be considered as a purely statistical alternative to a dynamical explanation 
of the neutron-rich IMF emission at midrapidity referring to the
"neck fragmentation" phenomenon \cite{dempsey,larochelle}. 
Theoretically such a process is an example of a new 
kind of statistical emission induced by an inhomogeneous external 
long--range field \cite{botvina99}.


\begin{thebibliography}{99}
\bibitem{PR95} J.P.Bondorf et al., {\em Phys. Rep.} {\bf 257}, 133 (1995).
\bibitem{mc_prc} A.S.Botvina and I.N.Mishustin. GSI preprint 2000-50, 
Darmstadt, 2000; nucl-th/0011072, 2000. 
\bibitem{msu} H.S.Xu et al., {\em Phys. Rev. Lett.} {\bf 85}, 716 (2000).
\bibitem{mueller} H.M\"uller and B.D.Serot, {\em Phys. Rev.} {\bf C52}, 2072 
(1995).
\bibitem{milazzo} P.M.Milazzo et al., {\em Phys. Rev.} 
{\bf C62}, 041602(R) (2000). 
\bibitem{botvina99} A.S.Botvina et al., {\em Phys. Rev.} {\bf C59}, 3444 (1999). 
\bibitem{dempsey} J.F.Dempsey {\it et al.}, {\em Phys. Rev.} {\bf C54}, 1710 (1996). 
\bibitem{larochelle} Y.Larochelle {\it et al.}, {\em Phys. Rev.} {\bf C62}, 
051602(R) (2000). 
\end{thebibliography}
\end{document}